\documentstyle[12pt,twoside,fleqn,espcrc1,epsfig,rotate]{article}

\newcommand{\AmS}{{\protect\the\textfont2
  A\kern-.1667em\lower.5ex\hbox{M}\kern-.125emS}}
\newcommand{\lapprox}{\raisebox{-0.5ex}{$\ 
\stackrel{\textstyle<}{\textstyle\sim}\ $}}

\title{Lessons from the 3d $U(1)$ 
Gross-Neveu Model}

\author{Susan E. Morrison \address{Department of Physics, University of Wales 
Swansea,\\Singleton Park, Swansea, SA2 8PP, U.K.}%
        \thanks{This work was supported by EU contract no. ERBFMRXCT97-0122}}

\begin{document}
\maketitle

\begin{abstract}
The effectiveness of the Glasgow algorithm is explored via implementation
in the 3d $U(1)$ Gross-Neveu model and the realisation of the Goldstone mechanism 
in this model is compared and contrasted with its realisation in QCD. 
\end{abstract}

\section{Introduction}
 In QCD the fermion determinant is complex for chemical potential, $\mu$
 non-zero, therefore generating an ensemble at $\mu \neq 0$  is
impracticable. The fermion 
number density, $J_0$, measures the excess of quarks relative to anti-quarks
and is expected to start to rise from zero at some value, $\mu_o$. The 
{\it onset}, $\mu_o$ corresponds to the point where the
phase of nuclear matter is more energetically favourable than the vacuum
state (for which $J_0=0$). Chiral symmetry restoration
occurs at $\mu_c$ and we expect \cite{MISHA2} that $\mu_0\lapprox\mu_c\simeq m_p/3$,
where $m_p$ is the proton mass. Quenched simulations, on the other hand,
predict $\mu_c \simeq m_{\pi}/2$ suggesting that in the
limit where the bare quark mass, $m_q\rightarrow 0$ chiral symmetry is
restored for any $\mu\neq0$. This result is unphysical because the
pion should not couple to the baryon chemical potential.

First attempts to simulate full QCD using
the Glasgow method \cite{BARBELL} also
produced perplexing results \cite{LAT96}.  
On an $8^4$ lattice with bare quark mass $m_q=0.01$ at $\beta=5.1$ we found  
$\mu_o \simeq 0.1$ which differs considerably from the strong coupling analysis \cite{BILIC}
prediction $\mu_c\simeq 0.65$ for $\beta=5.0$. Furthermore 
the scaling of $\mu_o$ with $m_q$ was consistent with a Goldstone
boson controlling the onset.
 
This result has
motivated an assessment of the effectiveness of the Glasgow method
via its implementation in a simpler model. The method involves expansion 
of the Grand Canonical Partition Function (GCPF) as a polynomial in the fugacity 
variable ($ e^{\mu / T}$). 

Consider the conventional expression for the GCPF in lattice QCD
\begin{equation}
Z\sim\int[dU]\,\det\left(M(U,\mu_{meas},m_q)\right)\,e^{-S_g(U)}
\end{equation}
The GCPF (for fixed $m_q$) can be rescaled and expressed as an ensemble average of $\det M$ at $\mu=0$: 
\begin{eqnarray}
Z(\mu) = 
{{ 
\int [dU] \, 
{{\det M(\mu_{meas})}\over{\det M(\mu_{upd}=0)}} 
\det M(\mu_{upd}=0)\,e^{-S_g[U]}
}
\over 
{
\int {[dU]\,\det {M(\mu_{upd}=0)\,e^{-S_g[U]}}}
}}\nonumber 
= \left<
{{\det M(\mu_{meas})} \over {\det M(\mu_{upd}=0)}} 
\right >{\biggr\vert}_{\mu_{upd}=0} 
\label{eqn:reweight}
\end{eqnarray}
where $\mu_{upd}$ is the chemical potential at which the statistical ensemble
is updated and is distinct from $\mu_{meas}$ which appears in 
the functional measure for the exact hybrid Monte Carlo (HMC) simulations and 
in $R_{rw}$ below.
 Note that generating the ensemble at $\mu_{upd}=0$ allows us to circumvent the
problem of the complex action in the HMC algorithm. For optimum efficiency of 
the Glasgow method we require 
a large overlap between the ensemble generated using $\det M (\mu_{upd})$ and the
exact ensemble generated using $\det M (\mu_{meas})$. Let us define a reweighting 
factor $R_{rw}\equiv{{\det M(\mu_{meas},m)} \over {\det M(\mu_{upd},m)}}$.
The relative magnitude of the factor $R_{rw}$ configuration by configuration
gives a measure of the overlap. If there is poor overlap between the simulated ensemble 
and the true ensemble it is conceivable that only a small fraction of the 
configurations will contribute significantly to $Z$ (those where
$R_{rw}$ is large in magnitude) in which case very
high statistics would be required to extract realistic observables.


The relevant features of the 3d Gross-Neveu (GN) model for $\mu\neq 0$ studies 
are that it has a chiral transition with a massless pion in the broken phase
and it can be formulated such that the fermion determinant is 
positive definite for $\mu \neq 0$.
 
  Mirroring the Glasgow reweighting technique implemented in
full QCD ($\mu \neq 0$) simulations we performed an expansion of the Grand
Canonical partition function (GCPF) for the 3d GN action in the fugacity variable 
$e^{\mu/T}$.

The full lattice action for the bosonized GN model with U(1) chiral symmetry 
is given in \cite{HKK3,SJHproc}. The functional measure used in the HMC algorithm is $\det(M^{\dagger}M)$.
The Dirac fermion matrices, $M$ and ${M}^{\dagger}$, can be 
conveniently expressed in terms of matrices $G$ and $V$ where $G$ contains
all the spacelike links while $V$ ($V^{\dagger}$) contains the forward(backward) timelike links 
\begin{equation}
       2iM_{xy}(\mu)=Y_{xy} + G_{xy} + V_{xy} e^{\mu} +V^{\dagger}_{xy} e^{-\mu} 
 \nonumber \;\;;\;\;
-2iM^{\dagger}_{xy}(\mu)=Y_{xy}^{\dagger} + G_{xy} + V_{xy} e^{\mu} +V_{xy}^{\dagger} e^{-\mu} \nonumber
\end{equation}
and the term describing the Yukawa couplings of scalars 
to fermions is given (in terms of the auxiliary fields $\sigma$
and $\pi$ on dual lattice sites $\tilde x$) by
\begin{equation}
Y_{xy} = 2i( m_q + \frac{1}{8} \sum_{<x,\tilde x>} \left(\sigma( \tilde x)
                   +i\epsilon\pi( \tilde x)\right))\delta_{xy}. 
\end{equation}

The determinants of these fermion matrices are related to that of
the propagator matrix P (following Gibbs \cite{GIBBS2}):
\begin{equation}
P=\left(\begin{array}{cc}
-GV-YV & V \\
  -V   & 0
\end{array} \right)
\end{equation}
by
\begin{equation}
\det(2iM) = e^{ \mu n_s^3 n_t} \det(P-e^{-\mu}) \;\;;\;\;
\det(2iM^{\dagger})= e^{\mu n_s^3 n_t} \det( (P^{-1})^{\dagger} - e^{-\mu})
\end{equation}

Since $\det M$ has been expressed in terms of the determinant of a matrix 
which is diagonal in $e^{-\mu}$ we can expand $\det M$ as a polynomial
in $e^{\mu}$.  
We can measure the averaged characteristic
polynomial over the ensemble generated at $\mu=0$ and provided that the 
coefficients are sufficiently well determined, we can use this to provide
an analytic continuation to any non-zero $\mu$. Consider a lattice with
$n_s$ spatial and $n_t$ temporal dimensions.
Determination of the eigenvalues of $P^{n_t}$ allows us to construct
the complete fugacity expansion for the GCPF:
\begin{equation}
{Z} = \sum_{n=-2 {n_s}^2 }^{2 {n_s}^2}\langle b_{|n|}\rangle e^{n\mu {n_t}} 
 = \sum_{n=-2 {n_s}^2}^{2 {n_s}^2}e^{-(\epsilon_n - n\mu)/T}.
\label{eqn:GCPF}
\end{equation}
 
  The expansion coefficients $\langle b_{|n|}\rangle$ are evaluated in the simulation
  and thermodynamic observables can be obtained from derivatives of $\ln Z$

The simulated 
3d GN $U(1)$ model has a positive definite functional measure 
so we can choose $\mu_{upd}\neq 0$ to investigate the influence 
of this choice on the observables. 
Exact hybrid Monte Carlo (HMC) simulations \cite{HKK3} showed a clear separation 
of the scales $m_\pi/2$ and $\mu_c$ indicating that the existence of a Goldstone
mode in the spectrum of a theory need not precipitate
chiral symmetry restoration for $\mu\ll m_{p}/3$ in QCD. 

Does the
poor overlap in the Glasgow algorithm prevent us from seeing the discontinuity in
the number density at $\mu_c\gg m_{\pi}/2$?

We will compare exact HMC and Glasgow method simulations on a $16^3$ lattice
at a four-fermi coupling of $1/g^2=0.5$ and $m=0.01$. 
We simulated the $N_f=12$ (N=3 staggered) model. In the exact
simulations for this parameter set there was a clear discontinuity 
(at $\mu_c=0.725(25)$) in
fermion number density as a function of the chemical potential. There 
was no evidence of an onset in the number density at 
$\mu \simeq m_{\pi}/2=0.18(1)$. In fact $\mu_o\lapprox \mu_c$
for the exact HMC simulation. 

\begin{figure}
\epsfig{file=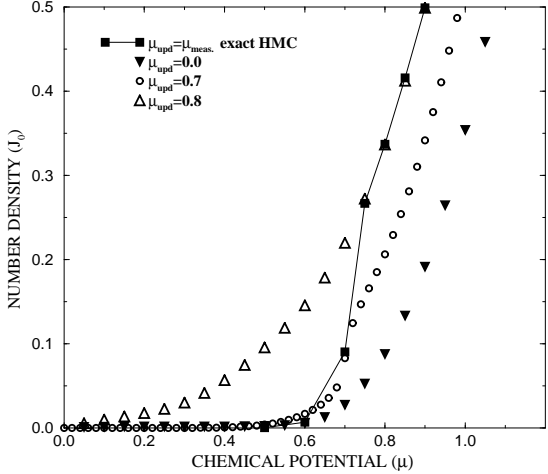,width=6.5cm,angle=-90,clip=}
\vspace{-0.3cm}
\caption{Fermion number densities comparing simulations for three different 
values of $\mu_{upd}$ with the exact hybrid Monte-Carlo simulation
on a $16^3$ lattice with $1/g^{2} =0.5$ and $m=0.01$.}
\label{fig:all_updates}
\end{figure}

It is clear that the chemical potential $\mu_{upd}$ at which the
statistical ensemble is generated has a 
strong influence on the thermodynamic observables of the simulation. 
Fig. \ref{fig:all_updates} shows the number density for the exact HMC simulation
(discussed above) and for three simulations with the Glasgow method:
one for $\mu_{upd}=0.0$ another for $\mu_{upd}=0.7(\lapprox \mu_c)$ 
and finally with $\mu_{upd}=0.8(>\mu_c)$  
The discontinuity at $\mu_c$ associated with the fermion losing
dynamical mass, which is clearly evident in the exact HMC data is not
consistently reproduced by the Glasgow algorithm. In fact the chiral transition
is seen only when $\mu_{upd}\lapprox\mu_{c}$.
For $\mu_{upd}>\mu_c$, $J_0$ reflects only the chirally symmetric phase 
(where $m_f=m_q$)
of the exact HMC data  while for $\mu_{upd}=0$, $J_0$ reflects only 
the phase of broken chiral symmetry (where $m_f\gg m_q$).
This suggests that there is insufficient overlap.
\begin{figure}
\vspace{-0.3cm}
\epsfig{file=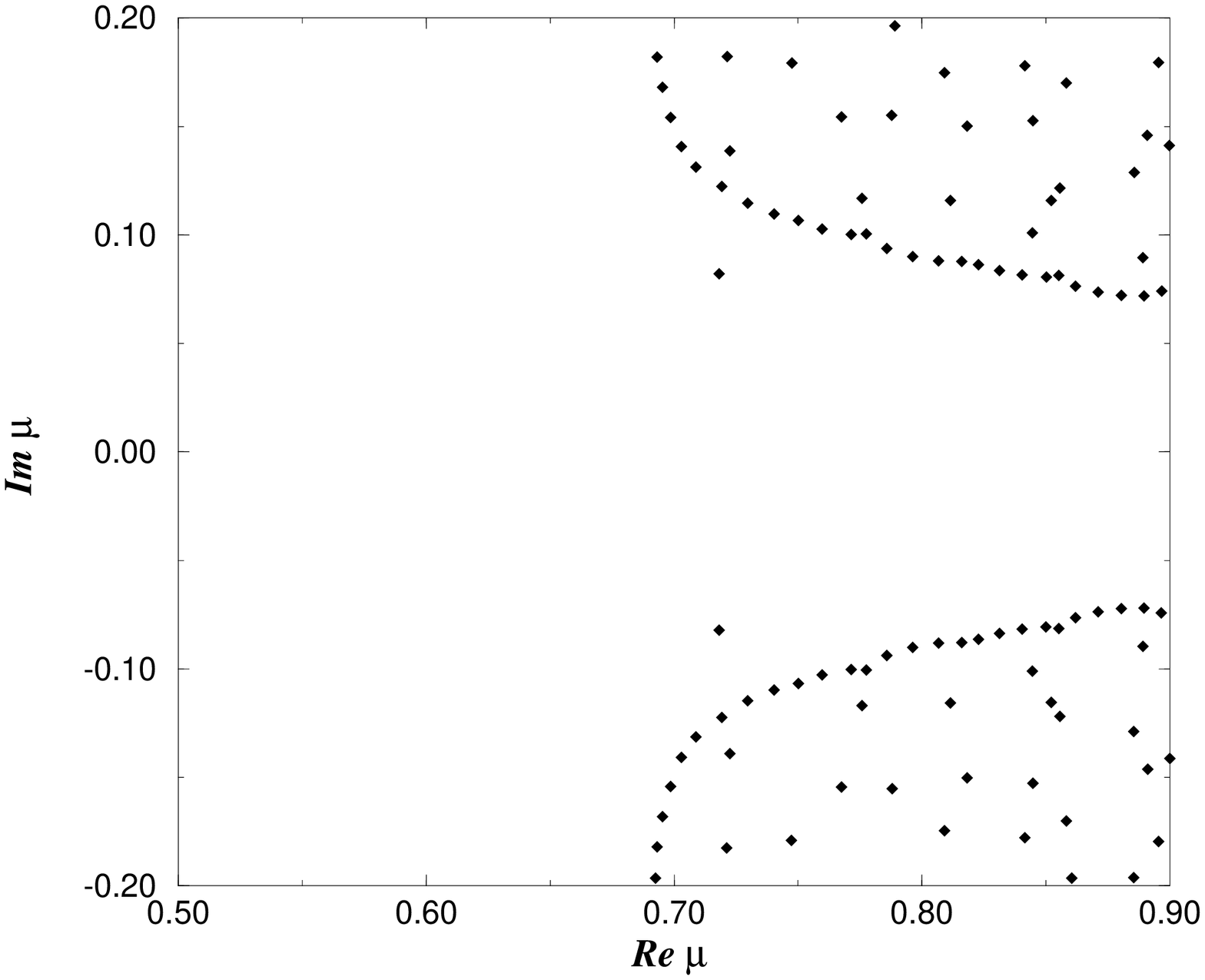,width=6cm,angle=-90,clip=}
\vspace{-0.5cm}
\epsfig{file=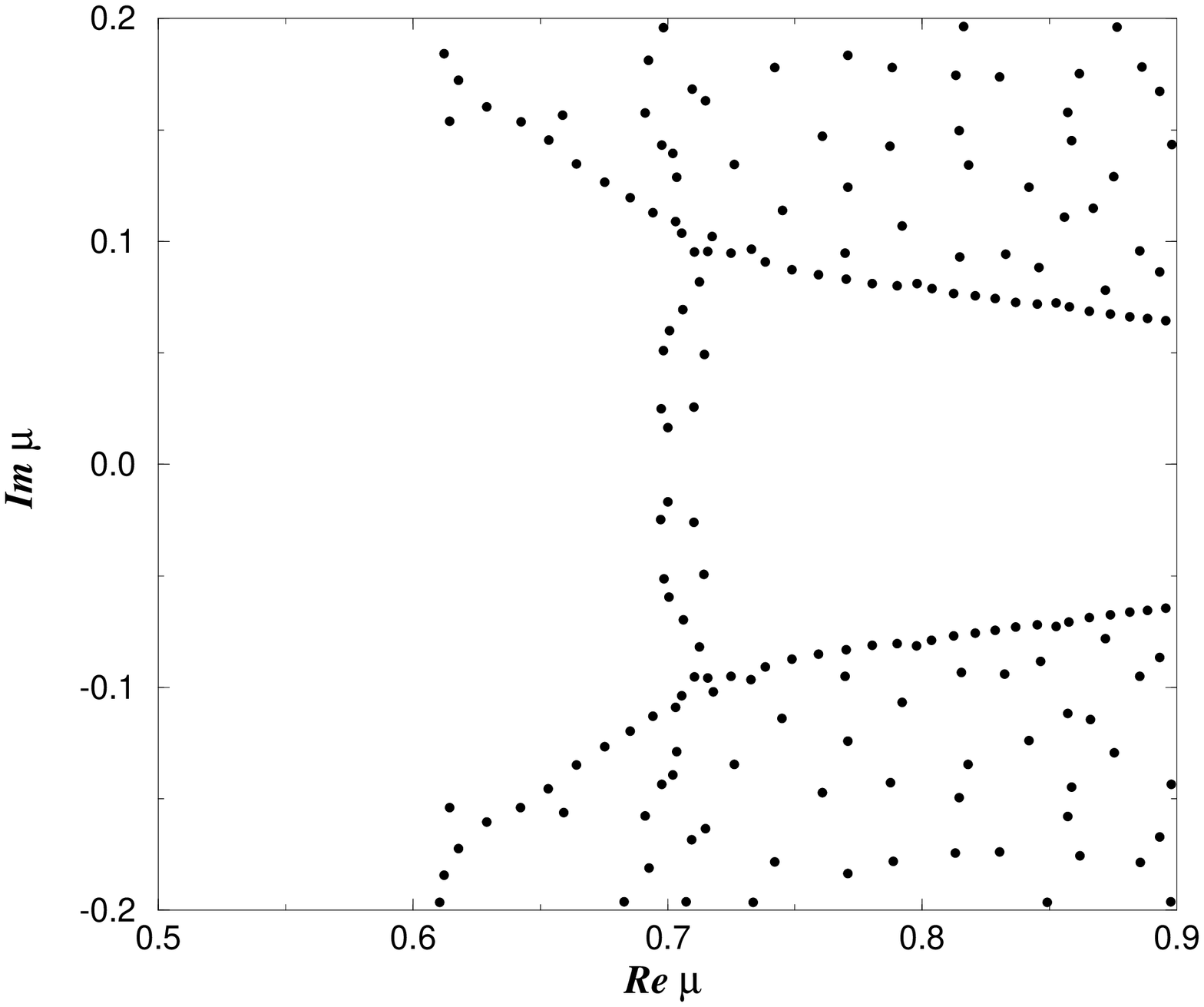,width=6cm,angle=-90,clip=}
\vspace{-0.2cm}
\caption{Partition function zeros for two simulations: upper
diagram with $\mu_{upd}=0.0$ showing two isolated zeros emerging from lobes
and lower diagram with $\mu_{upd}=0.7$ showing an arc of zeros approaching
the real axis at $\mu \simeq \mu_c$. Simulations on $16^3$ lattice with  
$1/g^{2} =0.5$ and $m=0.01$.}
\label{fig:zeros_0.0_0.7}
\end{figure}

The Lee-Yang zeros \cite{LEEYANG} in the complex $\mu$ plane are the zeros of 
Eqn. \ref{eqn:GCPF} and their distribution should reflect $\mu_c$. We expect
the zero with the smallest imaginary part to approach the real axis as
the lattice volume is increased. A phase transition occurs whenever a root 
approaches the real axis in the infinite volume limit.
The zeros for the $\mu_{upd}=0.0,0.7$  are plotted in 
Fig.\,\ref{fig:zeros_0.0_0.7}. For $\mu_{upd}=0$ notice
the two zeros 
emerging from the body of the distribution. These two isolated zeros are
located at a chemical potential $\mu \simeq \mu_c$. There was no evidence 
for $\mu_c$ in $J_0$ for $\mu_{upd}=0$ so it is more likely 
these two zeros are associated with $\mu_o$ rather than $\mu_c$.
For $\mu_{upd}=0.7$ we {\bf did} see a discontinuity in $J_0$ at
$\mu_c$ and in this case we see an arc of zeros forming which intersects
the real $\mu$-axis at $\mu \simeq \mu_c$.


How do we explain the fact that $\mu_o \simeq \mu_c$ in simulations of this model but
not in QCD?
Consider the lattice Ward identity for the chiral condensate: 

\begin{eqnarray}
\sum_{y} \langle {\bar{\psi}\gamma_{5}\psi(y)\bar{\psi}
\gamma_{5}\psi(x)}\rangle 
&=&\sum_{y}\langle{tr(G_{-\mu}^{\dagger}(x,y)G_{+\mu}(x,y))}\rangle-
\langle {(tr\gamma_{5}G(x,x))(tr\gamma_{5}G(y,y))}\rangle  \nonumber \\
&=& -{\frac{\langle {\bar{\psi}(x)\psi(x)}\rangle}{m_q}}
\label{eqn:ward}
\end{eqnarray}

The pion susceptibility Eqn.(\ref{eqn:ward})
consists of a connected channel and a disconnected channel.
In QCD the pion has a {\bf dominant connected contribution}
therefore  we can identify the pion mass in QCD with a pole
in the pseudoscalar propagator, $G_{ps}$, defined by (with $G=M^{-1}$):
\begin{equation}
G_{ps}(t)=\sum_{\vec{x}} G_{+\mu}(\vec{x},t)G_{-\mu}^{\dagger}(\vec{x},t) \simeq e^{-m_{\pi}t}
\end{equation}

Gibbs \cite{GIBBS2} derived a relation between the eigenvalues of $P$
and $m_\pi$ in QCD. If $\lambda_{ps}$ is an eigenvalue of P associated with 
the mass pole in $G_{ps}$ and assuming the pion susceptibility has a dominant 
connected contribution it follows that $m_{\pi} = 2 \ln |\lambda_{ps}|$.
Let us consider how the existence of this mass pole could affect $\mu_{o}$.
Notice that $\det M$ can be simply expressed in terms of the eigenvalues of $P$
and $Z$ can be similarly expressed in terms of its zeros $\alpha_i$ in the $e^{\mu}$ plane:
\begin{equation}
\det M = e^{n_{s}^3n_{t} \mu} \prod_{i=1}^{4n_{s}^3n_{t}}(e^{\mu}-\lambda_i) \;\;\;;\;\;\;
Z = e^{ n_{s}^3n_{t} \mu} \prod_{i=1}^{4n_{s}^3n_{t}}(e^{\mu}-\alpha_i).
\end{equation}

Since $Z = \langle \det (M)\rangle$ we see that on a single configuration 
$\alpha_i=\lambda_i$. Note however that the ensemble
averaged $\alpha_i$'s are not in general the same as the ensemble 
averages of the $\lambda_i$'s.  

The number density on a single configuration 
$J_0^i\sim {{\partial \ln \det M}/{\partial \mu}}$ whereas the ensemble
average is given by $J_0\sim {{\partial \ln \langle\det M\rangle}/{\partial \mu}}$.
Presumably the unphysical singularity at $\mu_o\simeq m_{\pi}/2$ in 
$J_0$ associated with the pole in $G_{ps}$ at 
$m_\pi/2$ could disappear if we achieved the $\alpha_i$ appropriate to the
correct statistical ensemble and we envisage that $\mu_o$ would cancel in 
the large ensemble limit. The mass pole 
at $m_{\pi}/2$ should disappear if we have a confining theory.

It will be shown that in the 3d GN $U(1)$ model the Goldstone pole forms in the 
disconnected channel therefore the state described by $G_{ps}$
no longer corresponds to the Goldstone pion. Instead we find:
\begin{equation}
G_{ps} \simeq e^{-2m_f(\mu=0)t}
\end{equation}
Since $|\lambda_{ps}|$  will now correspond to the dynamical fermion
mass $m_f$ rather than $m_{\pi}/2$ we have no reason 
to expect an early onset associated with the Goldstone pole in the
3d GN $U(1)$ model. 
\begin{figure}
\vspace{-0.3cm}
\epsfig{file=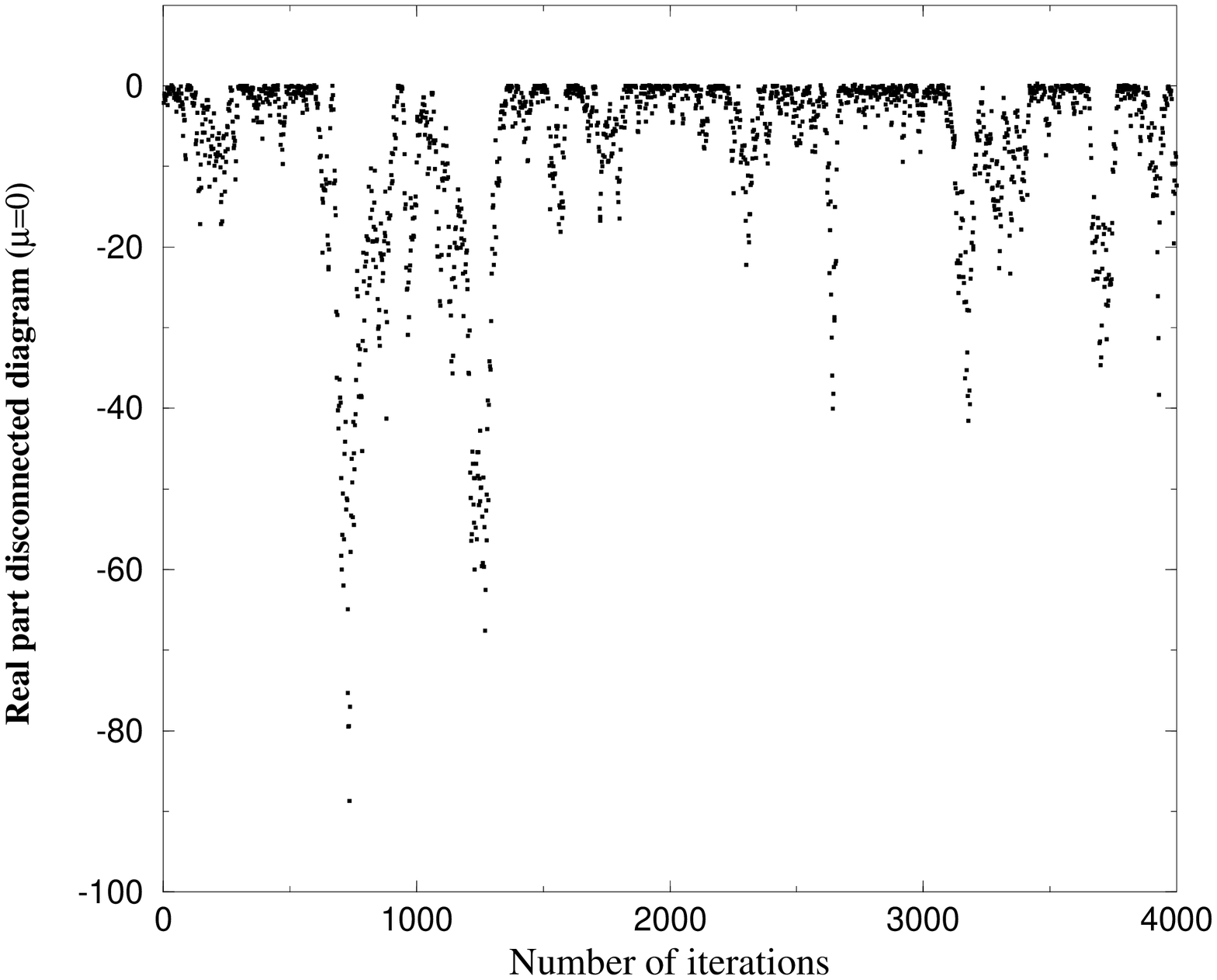,width=6cm,angle=-90,clip=}
\vspace{-0.3cm}
\epsfig{file=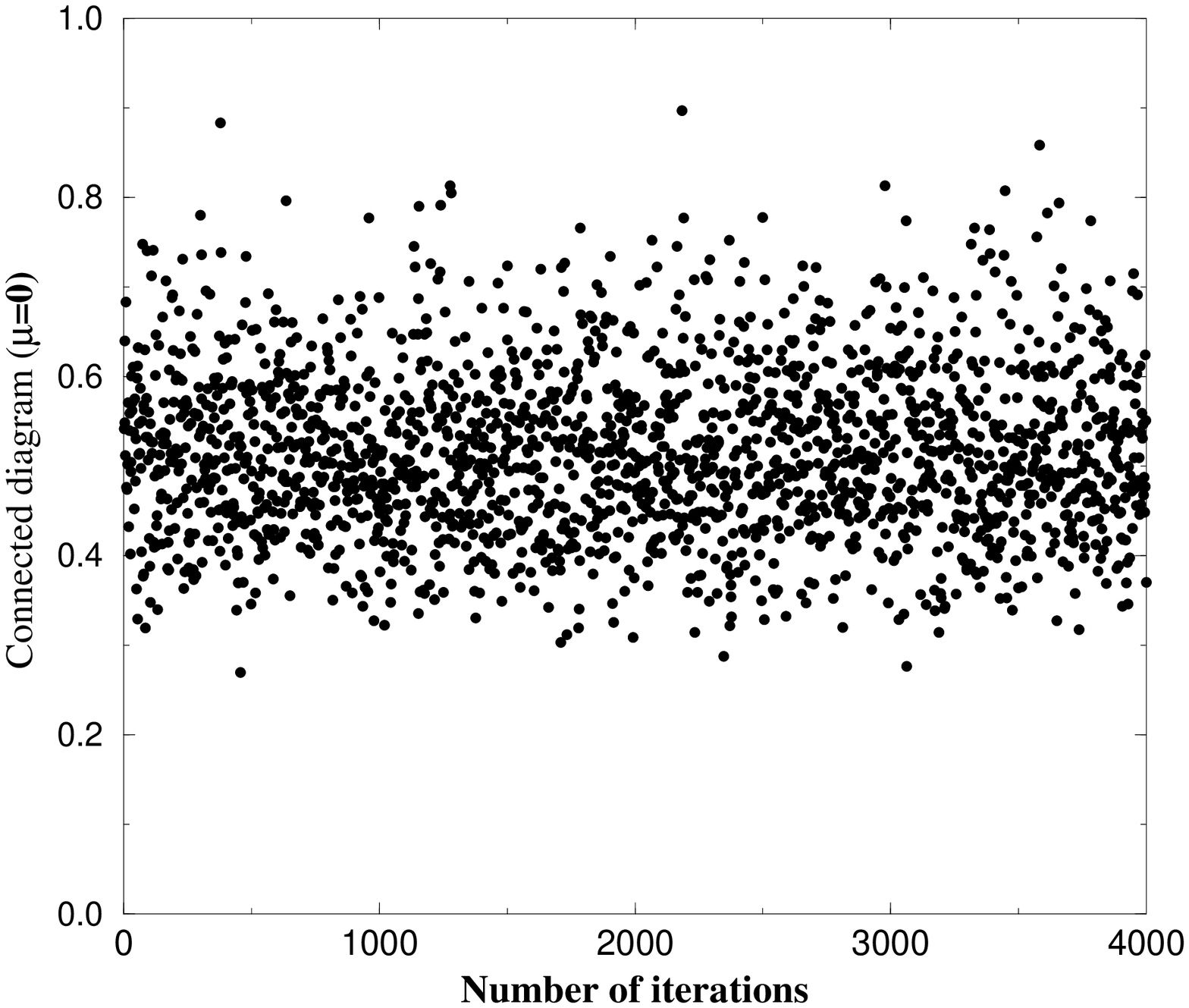,width=6cm,angle=-90,clip=}
\caption{ Measurements at $\mu=0.0$  showing the disconnected 
and connected contributions 
to the pion susceptibility on a $16^3$ lattice with $1/g^{2} =0.5$ and 
$m=0.01$.}
\label{fig:mu0discon_pisq}
\end{figure}  

 Fig. \ref{fig:mu0discon_pisq} shows the disconnected
contribution to the Ward identity for a simulation at
zero chemical potential. For this simulation we found
${{\langle\psi\bar{\psi}\rangle}/{m_q}} \simeq 40$, therefore we require that the
sum of the connected and disconnected diagrams be of similar order
so that Eqn. \ref{eqn:ward} is satisfied. The connected contribution was 
stable at a value of around 0.5 therefore the dominant contribution
must come from the disconnected diagram. We found that the disconnected
contribution  was very noisy with large
downward peaks. The data suggests that considering
the connected contribution alone will never be sufficient to satisfy 
Eqn. \ref{eqn:ward}.  

 We repeated our measurements for
a non-zero chemical potential. We chose $\mu=0.5$ thus ensuring that we 
were still in the phase of broken chiral symmetry.
The results were consistent with those at zero chemical potential
as one would expect.

The disconnected contribution to the pion susceptibility
gives a very noisy signal which suggests 
that a very long run would be required to equilibrate sufficiently 
to satisfy the lattice Ward identity. 


The Glasgow algorithm is most 
effective at predicting $\mu_c$ for $\mu \lapprox \mu_c$ when the 
configurations of the statistical ensemble reflect the system close to criticality .

 In QCD the pseudoscalar channel pole is formed from {\bf connected}
 diagrams corresponding to $G_{+\mu}(t)G^{\dagger}_{-\mu}(t)$. A 
 {\bf baryonic pion} forms from a quark and a conjugate quark and condenses 
 in this channel which explains
 why we find $\mu_{o}=m_{\pi}/2$ \cite{OURNEW}. This induces the unphysical early onset of chiral 
 symmetry restoration in the quenched theory and will persist in the Glasgow
 algorithm unless either large statistics or sufficient overlap is achieved
 so that the phase of the determinant eliminates the conjugate quarks.  

In 3d GN $U(1)$ the Goldstone mechanism is realised by a
pseudoscalar channel pole formed from {\bf disconnected} diagrams and 
the state  $G_{+\mu}(t)G^{\dagger}_{-\mu}(t)$ yields a bound state
of mass $2m_f(\mu=0)$ which is considerably heavier than the pion. Even on 
individual configurations in this model we expect $\mu_o\simeq\mu_c\gg m_{\pi}/2$
because no baryonic pion condenses in the connected channel.

\end{document}